# Artificial Multiferroics and Enhanced Magnetoelectric Effect in van der Waals Heterostructures

*Yan Lu,[1,2] Ruixiang Fei,[1] Xiaobo Lu,[1] Linghan Zhu,[1] Li Wang,[2]\* and Li Yang[1]\**

[1]Department of Physics and Institute of Materials Science and Engineering, Washington University, St. Louis, MO 63130, USA
[2]Department of Physics, Nanchang University, Nanchang 330031, China

**Abstract:** Multiferroic materials with coupled ferroelectric and ferromagnetic properties are important for multifunctional devices due to their potential ability of controlling magnetism via electric field, and vice versa. The recent discoveries of two-dimensional ferromagnetic and ferroelectric materials have ignited tremendous research interest and aroused hope to search for two-dimensional multiferroics. However, intrinsic two-dimensional multiferroic materials and, particularly, those with strong magnetoelectric couplings are still rare to date. In this paper, using first-principles simulations, we propose artificial two-dimensional multiferroics via a van der Waals heterostructure formed by ferromagnetic bilayer chromium triiodide ($CrI_3$) and ferroelectric monolayer $Sc_2CO_2$. In addition to the coexistence of ferromagnetism and ferroelectricity, our calculations show that, by switching the electric polarization of $Sc_2CO_2$, we can tune the interlayer magnetic couplings of bilayer $CrI_3$ between ferromagnetic and antiferromagnetic states. We further reveal that such a strong magnetoelectric effect is from a dramatic change of the band alignment induced by the strong build-in electric polarization in $Sc_2CO_2$ and the subsequent change of the interlayer magnetic coupling of bilayer $CrI_3$. These artificial multiferroics and enhanced magnetoelectric effect give rise to realizing multifunctional nanoelectronics by van der Waals heterostructures.

# 1. INTRODUCTION

Coupled ferroelectricity and ferromagnetism, *i.e.*, the magnetoelectric effect, play critical roles for applications of multifucntional devices due to its unique ability of tuning magnetism by applying electric field, and vice versa.[1,2] However, multiferroic materials with the strong magnetoelectric effect are still scarce due to a few well-known challenges. First, ferroelectricity is usually from the *s* and *p* electrons and requires materials with empty or fully occupied *d* orbitals. However, ferromagnetism is usually from *d* electrons and requires those with partially occupied *d* orbitals.[3] Second, the coupling between ferroelectricity and ferromagnetism is usually weak for multiferroic materials, in which ferroelectricity and ferromagnetism come from two types of *d*-orbital elements.

Recent advances in realizing ferroics in two-dimensional (2D) materials may give hope to overcoming these fundamental challenges of traditional bulk materials. For example, novel 2D materials, such as group-IV monochalcogenides,[4–8] group-V monolayer,[9] $III_2$-$VI_3$ van der Waals (vdW) materials,[10–14] transition metal thiophosphate (TMTP) family,[15–18] and oxygen-functionalized scandium carbide MXene ($Sc_2CO_2$),[19] are reported to be ferroelectric (FE) materials because their vdW interfaces can avoid the known depolarization effect of bulk structures.[20–22] Meanwhile, 2D ferromagnetism has been recently achieved in vdW materials, such as few-layer $Cr_2Ge_2Te_3$,[23] $CrI_3$,[24] and $Fe_3GeTe_2$,[25,26] spurring substantial interests to fabricate ultra-thin magnetic devices and study fundamental properties, such as magnons, spin liquids, and many other quantum states in reduced-dimensional structures.[27–44] All these advances pave the way for exploring the existence of 2D multiferroics and magnetoelectronic couplings.

To date, there have been predictions of 2D multiferroic materials. Monolayer group-IV monochalcogenides, e.g., SnSe and GeSe, have been predicted to be multiferroic with the coexistence of ferroelasticity and ferroelectricity.[4,45] Hole-doped $\alpha$-SnO monolayer has been predicted to be multiferroic with the coexistence of ferromagnetism and ferroelasticity.[46] Monolayer TMPCs-$CuMP_2X_6$ (M=Cr, V; X=S, Se)[47] and charged $CrBr_3$ monolayer[48] have been predicted to exhibit multiferroicity with the coexistence of ferromagnetism and ferroelectricity. However, these structures do not exhibit strong magnetoelectronic effects because of the intrinsic limit of the weak coupling between different $d$-orbital elements. Interestingly, a recent work by Zhang' group[49] predicted that multiferroics can be achieved by artificial vdW heterostructures, and the magnetic easy axis of $Cr_2Ge_2Te_3$ can be tuned by the polarization of the nearby ferroelectric layer. These advance shed light on searching for multiferroics and further turning on/off magnetism in artificial vdW heterostructures.

In this work, using first-principles density functional theory (DFT) simulations, we propose 2D artificial multiferroic realized by a vdW heterostructure via stacking bilayer $CrI_3$ and monolayer $Sc_2CO_2$. Importantly, we show that tuning the polarization direction of the $Sc_2CO_2$ layer can switch the interlayer magnetic order of bilayer $CrI_3$ to be ferromagnetic (FM) or antiferromagnetic (AFM), *i.e.*, turning on/off the magnetism. Our analysis indicates that this enhanced coupling between ferroelectricity and magnetic order is from the significant change of interlayer band alignment and subsequent interlayer charge transfers. Therefore, artificial vdW multiferroics may provide ample space to realize strong magnetoelectric couplings for future multifuncational nanoelectronics and devices.

## 2. RESULTS AND DISCUSSION

The 2D artificial multiferroics is constructed by stacking bilayer $CrI_3$ with monolayer $Sc_2CO_2$ to form a vdw 2L-$CrI_3$/$Sc_2CO_2$ heterostructure. The in-plane lattice constant of free-standing $CrI_3$ is 6.87 Å, and that of $Sc_2CO_2$ is 3.37 Å. By using a 2×2 supercell of monolayer $Sc_2CO_2$ to match bilayer $CrI_3$, as shown in Figure 1a, the in-plane lattice mismatch can be reduced to be less than 2%, which is widely accepted in first-principles studies.[49–51]

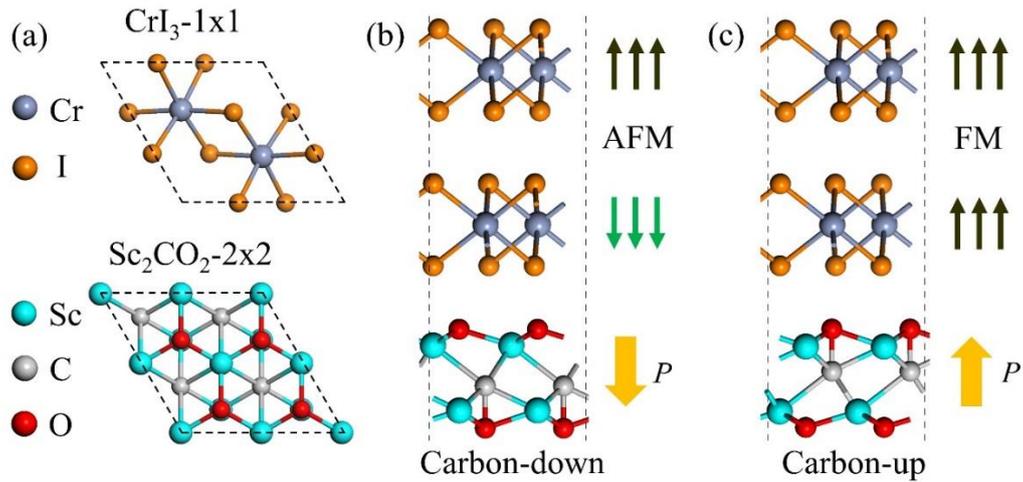

**Figure 1.** (a) Top-views of 1×1 unit cell of $CrI_3$ and 2×2 unit cell of $Sc_2CO_2$. (b) and (c) Side views of 2L-$CrI_3$/$Sc_2CO_2$ heterostructures with carbon-down and carbon-up configurations of the $Sc_2CO_2$ layer, respectively. The black and green thin arrows denote the spin directions in 2L-$CrI_3$. The golden thick arrow denotes the direction of the out-of-plane electric polarization in $Sc_2CO_2$.

Monolayer $Sc_2CO_2$ is predicted to exhibit both in-plane and out-of-plane electric polarizations.[19] Free-standing monolayer $Sc_2CO_2$ exhibits an out-of-plane electric polarization of 1.60 $\mu C/cm^3$, which is reported to be the largest for a monolayer material. This polarization can be switched by the displacement of the C atoms, as denoted by carbon-down and carbon-up configurations shown in Figures 1b and 1c. The calculated energy barrier is 0.52 eV per formula unit between the Carbon-down and Carbon-up configurations for $Sc_2CO_2$.[19] The coercive field can be up to 2.5 V/nm if the temperature effect is not taken into account. Howver, this coercive field can be substantially reduced after condiering finite temperature.

CrI$_3$ is an intensively studied 2D magnetic material. With the help of out-of-plane magnetic anisotropy,[52] the 2D structure hold a long-range magnetic order by gapping low-energy modes of magnons. It has two types of interlayer stacking styles: the high-temperature (HT) and low-temperature (LT) phases.[27] Importantly, despite the robust intralayer FM order, the interlayer magnetic coupling of few-layer CrI$_3$ is sensitive to the stacking structure: it is FM for the LT phase but AFM for the HT phase.[53–56] Surprisingly, most experimental samples exhibit the HT phase with a subsequent layer-dependent magnetism.[35,57–61] Therefore, we choose the HT-phase of bilayer CrI$_3$ in our studied heterostructure. More thorough studies reveal that the energy difference between interlayer FM and AFM order is so small (~ 0.1 meV/ Cr ion), explain the highly tunable (by strain, doping, and magnetic field) magnetism observed in the HT-phase CrI$_3$. This tunable interlayer magnetism also gives hope to realizing magnetoelectric effects in this 2D structure.

After forming the 2L-CrI$_3$/Sc$_2$CO$_2$ heterostructure shown in Figure 1, we find that the FE order of the Sc$_2$CO$_2$ layer and the magnetic order of bilayer CrI$_3$ coexist, indicating the multiferroics. More interestingly, the out-of-plane electric polarization from the neighboring Sc$_2$CO$_2$ layer can substantially switch the magnetic order of bilayer CrI$_3$. As shown in Figure 1b, when the Sc$_2$CO$_2$ layer is in the carbon-down configuration with a downward polarization, bilayer CrI$_3$ keeps the AFM interlayer order, and the AFM state is about 0.37 meV/unit cell lower than the FM state. However, when the Sc$_2$CO$_2$ layer switches to the carbon-up configuration with an upward polarization

(Figure 1c), bilayer CrI$_3$ will be tuned to the FM interlayer order, and the FM state is about 1.41 meV/unit-cell lower than the AFM state. As a result, the overall magnetism can be turned on or off by the electric polarization (or external gate field), resulting in a significant magnetoelectric effect.

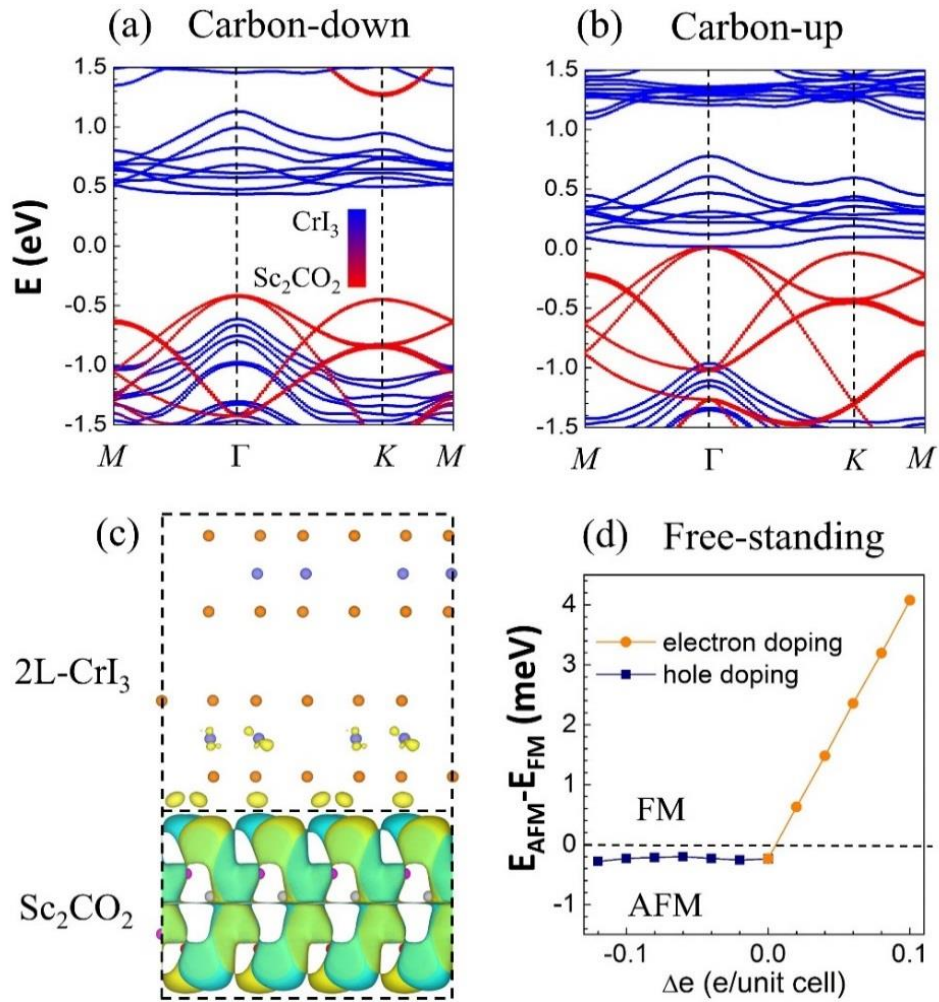

**Figure 2.** The DFT+U calculated band structure of 2L-CrI$_3$/Sc$_2$CO$_2$ heterostructures for (a) carbon-down and (b) carbon-up configurations, respectively. Colors indicate the weight of bands from CrI$_3$ (blue) and Sc$_2$CO$_2$ (red). (c) The differential charge distribution of the 2L-CrI$_3$/Sc$_2$CO$_2$ heterostructure calculated by subraction of the electron distribution of the carbon-down configuration from that of the carbon-up

configuration. The isosurface value is 0.001 $e$/Bohr$^3$. The yellow color means electron accumulation, and the cyan color means electron loss. (d) Electron and hole doping effects on the energy difference between AFM and FM interlayer couplings for free-standing bilayer HT CrI$_3$.

To understand such multiferroic and magnetoelectric coupling, we have plotted the electronic band structures of these 2L-CrI$_3$/Sc$_2$CO$_2$ heterostructures with the electric polarization up and down, respectively. In Figure 2a, for the carbon-down configuration, a finite band gap around 0.85 eV is observed, and the heterostructure forms a type II band alignment, in which the valence band maximum (VBM) comes from the Sc$_2$CO$_2$ layer, and the conduction band minimum (CBM) comes from bilayer CrI$_3$. However, for the carbon-up configuration, the band gap is closed under the polarization field. The CBM from bilayer CrI$_3$ touches the VBM from the Sc$_2$CO$_2$ layer, forming a type-III band alignment, as shown in Figure 2b.

Our further analysis shows that such a band crossing will introduce a charge transfer between CrI$_3$ and Sc$_2$CO$_2$ layers. In Figure 2c, we plot the variation of charge densities of the heterostructures under opposite polarizations, in which the electron distribution of the carbon-down configuration is substracted from that of the carbon-up configuration. In Figure 2c, despite complicated variations of the charge density in the Sc$_2$CO$_2$ layer due to the change of atomic positions related with ferroelectricity, we can see that there are extra electrons in the bilayer CrI$_3$ part of the carbon-up configuration, compared with that of the carbon-down configuration. According to the Bader's charge analysis, there are around 0.1 $e$ transferred from $Sc_2CO_2$ to $CrI_3$ in the carbon-up configuration, resulting in electron doping in $CrI_3$. On ther other hand, within the error bar of DFT calculations, we do not observe any significant charge transfer from $Sc_2CO_2$ to $CrI_3$ in the carbon-down configuration.

To further connect this charge transfer with the change of magnetism, we have checked the doping effect on free-standing bilayer $CrI_3$. As shown in Figure 2d, the HT-phase bilayer $CrI_3$ experiences a phase transition from the interlayer AFM to FM phases under electron doping but not hole doping, and the critical electron doping density is less than 0.01 $e$ per unit cell. In fact, such an electron-doping induced magnetic phase transition was observed in recent experiments of bilayer $CrI_3$.[41,42] Therefore, we conclude that the out-of-plane polarization of the $Sc_2CO_2$ layer dramatically changes the band alignment in the $CrI_3/Sc_2CO_2$ heterostructure, and the charge-transfer induced electron doping drives the magnetic phase transition of bilayer HT-phase $CrI_3$.

The energy difference between energy gaps of the FM and AFM orderings may be the reason for the electron-doping induced magnetic crossover. According to our calculations, the energy gap of FM 2L-$CrI_3$ is about 70-meV smaller than that of AFM 2L-$CrI_3$, which agrees with a previous calculation.[53] As a result, doped electrons will acquire less energy in the FM state than that in the AFM state. Giving the small energy difference between the interlayer FM and AFM orderings (~ 0.37 meV), for a reduction of 70 meV of band gap, a doping density of 0.01 $e$/unit-cell can provide about 0.7 meV energy gain of the AFM state than that of the FM state, making it possible to explain the energetic preference of the FM ground state. However, we have also noticed other mechanisms that can be applied to answering this question, such as the enhanced FM coupling mediated by interlayer electron hopping in itinerant electron systems.[62]

In our studied heterostructures, because of the broken symmetry and existence of polarization field, the doped charge is mainly accumulated at the bottom layer of $CrI_3$. On the other hand, in our calculations to explain doping effects in free-standing 2L-$CrI_3$ structures, the doped charge is evenly distributed to both bottom and top layers due to the preservation of symmetry. In this sense, these two simulations are not perfectly consistent with each other. However, we believe that this may quantitively affect our result but not fundamentally change the explanation and conclusion because the mechanism of the switch of magnetic ordering is from the band gap variation or itinerant electrons. Particularly, DFT may overestimate the inhomogeneous charge density under polarization field because it is known for not accurately capturing the screening effect. This is also evidenced by recent experiments, in which they gated and doped bilayer $CrI_3$ simultaneously.[41,42] In those experiments, doped charge is biased by the gating field but accumulated slightly difference between the top and bottom layers. More future works are valuable to quantitatively investigate the inhomogeneous charge distribution in these heterostructures.

However, DFT calculations are known for underestimating the band gap of semiconductors and insulators. Therefore, the above bandgap closing may be an artificial result of DFT calculations. To answer this concern, we have employed the Heyd–Scuseria–Ernzerhof (HSE) hybrid functional[63] to check the band alignment in the CrI$_3$/Sc$_2$CO$_2$ heterostructure. Because of the large number (284) of electrons in one unit cell of the 2L-CrI$_3$/Sc$_2$CO$_2$ heterostructure, it is formidable to perform brute-force HSE calculations with spin-orbit coupling (SOC) included. In the following, we will do two approximations to obtain the band alignments.

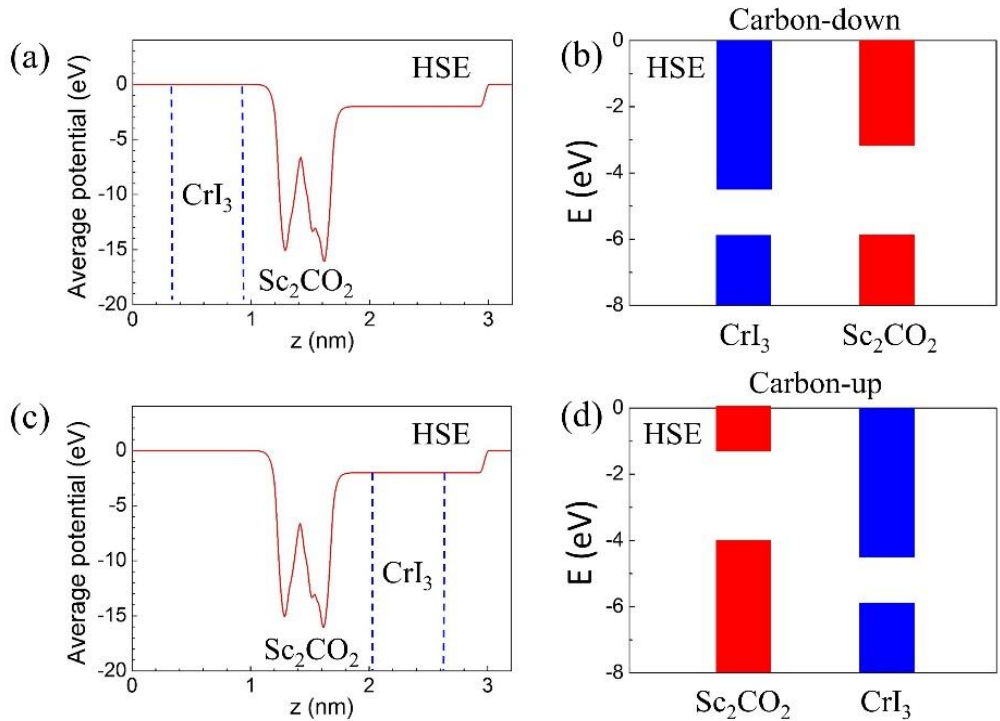

**Figure 3.** (a) Work function of Sc$_2$CO$_2$ calculated by HSE. (b) The band alignment of 2L-CrI$_3$/Sc$_2$CO$_2$ for the carbon-down configuration, in which CrI$_3$ is placed on the left side of Sc$_2$CO$_2$ as denoted by the two dashed lines in (a). (c) and (d) are similar to (a) and (b), but corresponding to the carbon-up configuration.

First, we perform the HSE calculations to obtain the work functions of isolated monolayer $Sc_2CO_2$ and 2L $CrI_3$, respectively. As shown in Figure 3a, the vacuums at the two sides of the $Sc_2CO_2$ layer have different potentials due to its spontaneous out-of-plane electric polarization. This potential difference is about 2.0 eV, which imply a very strong built-in electric field. As shown in Figure 3a, the $CrI_3$ bilayer will feel two different potentials depending on the polarization direction, inducing two types of the energy alignment for $CrI_3$ and $Sc_2CO_2$ heterostructures. By comparing these absolute band energies, we obtain a type-II band alignment in the carbon-down configuration and a type III band alignment in the carbon-up configuration in the 2L-$CrI_3$/$Sc_2CO_2$ heterostructures, as shown in Figures 3b and 3d, respectively. This agrees with the bandgap closing and charge transfer results for the carbon-up configuration in DFT calcualtions.

The above work-function calculations ignore the interlayer interactions, which may introduce error bars. Therefore, we try the second approach, in which we directly calculate the density of states (DOS) of the heterostructure but with one layer of $CrI_3$ and one layer of $Sc_2CO_2$ heterostructure by HSE. As shown in **Figures 4**a and 4b, the band structure and work functions of monolayer and bilayer $CrI_3$ are similar because the localized *d* orbitals aound band edges are not very sensitive to quantum confinement. Therefore, we can expect that the band alignment of the heterostructure made by monolayer $CrI_3$ and $Sc_2CO_2$ is similar to that made by bilayer $CrI_3$ and $Sc_2CO_2$. As shown in Figures 4c and 4d, we find that it is a type II band alignment for the carbon-down configuration and a type-III band alignment for the carbon-up configuration. Therefore, these HSE calculations convince our previous charge-transfer conclusions that the carbon-up configuration of $CrI_3/Sc_2CO_2$ is a type III metal state, which implies that the interlayer magnetic coupling of bilayer $CrI_3$ can also be changed from AFM to FM states when $Sc_2CO_2$ is changed from the carbon-down configuration to the carbon-up configuration.

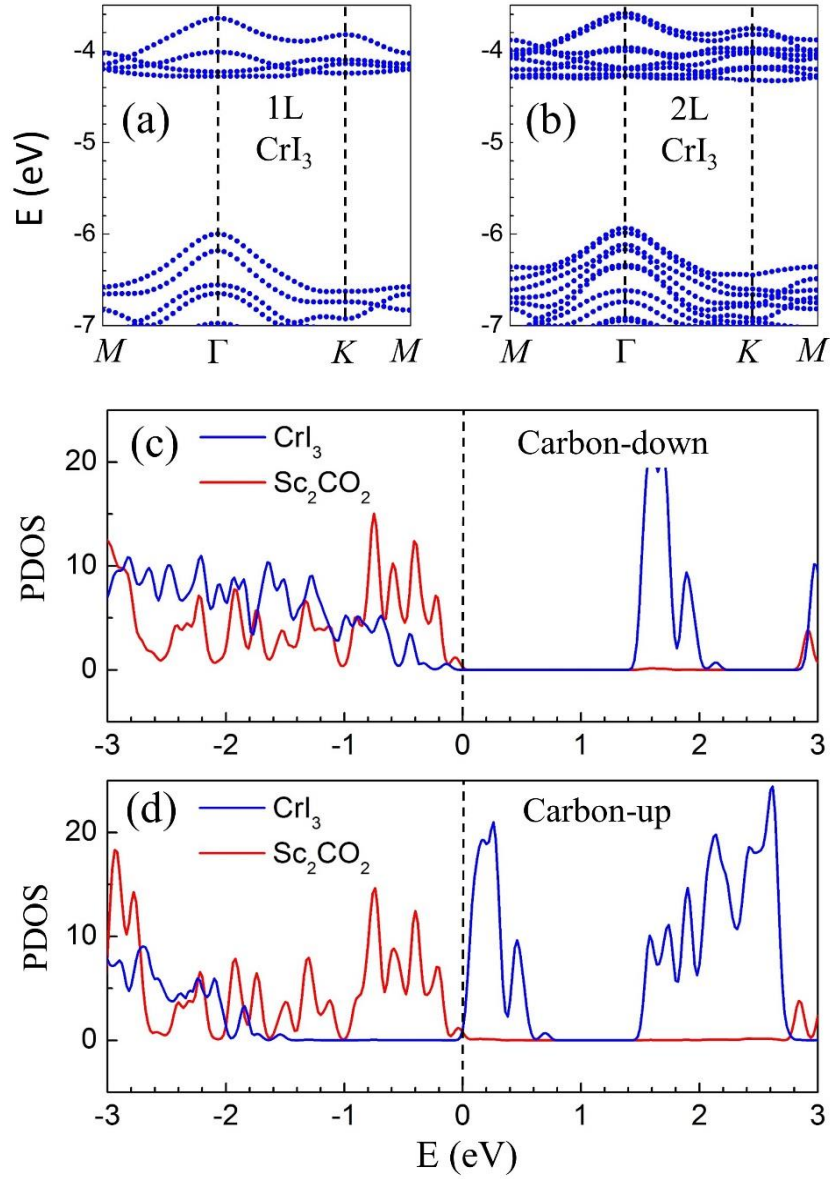

**Figure 4.** Band structures of (a) 1L-$CrI_3$ and (b) 2L-$CrI_3$ calculated by HSE. The HSE calculated PDOS of 1L-$CrI_3$/$Sc_2CO_2$ heterostructures with (c) carbon-down, and (d) carbon-up configurations. The blue and red curves are PDOS from $CrI_3$ and $Sc_2CO_2$, respectively.

**Table 1**. Energy difference between AFM and FM interlayer couplings of bilayer $CrI_3$ in the 2L-$CrI_3$/$Sc_2CO_2$ heterostructure calculated by different values of $U$ and $J$.

| $E_{AFM} - E_{FM}$ (meV/unit-cell) | $U$ = 3.9 eV $J$ = 1.1 eV | $U$ = 3.9 eV $J$ = 0.0 eV | $U$ = 2.9 eV $J$ = 0.7 eV | $U$ = 3.0 eV $J$ = 0.0 eV | $U$ = 3.9 eV, $J$ = 1.1eV (fully relaxed) |
|---|---|---|---|---|---|
| **Carbon-down** | -0.37 | -0.23 | -0.27 | -0.39 | -0.32 |
| **Carbon-up** | 1.41 | 3.41 | 2.27 | 2.6 | 1.60 |

It is known challenging for DFT simulations to describe the correlated effects in magnetic materials. Similar to most previous works in this field, we have employed the DFT+U approach to capture the localized $d$ orbitals. In this work, we choose the Liechtenstein scheme with $U$ = 3.9 eV and $J$ = 1.1 eV, which are calculated by the linear response method and have been successfully applied to explaining the observed AFM and FM interlayer magnetic orders of HT and LT phases of few-layer $CrI_3$.[53] On the other hand, to further confirm that our above conclusions are robust against the choices of $U$ and $J$ parameters, we have done the calculations with different values of $U$ and $J$ within a reasonable range, as used in previous calculations.[54–56] As summarized in **Table 1**, all these choices give the same conclusion that the AFM interlayer order is the ground state for the carbon-down configuration of $Sc_2CO_2$, and the FM interlayer order is the ground state of the carbon-up configuration of $Sc_2CO_2$.

It is interesting to check whether the magnetic easy axis of 2L-$CrI_3$ can be turned by ferroelectric states of $Sc_2CO_2$ since the switch of magnetic easy axis was predicted in $Cr_2Ge_2Te_3$ and charge doped CrSBr.[49,64] We have calculated the magnetic anisotropy energy (MAE), defined by the energy difference between magnetic states with out-of-plane and in-plane spins of 2L-$CrI_3$ under two ferroelectric states of $Sc_2CO_2$ in the heterostructure. The calculated results are shown in Table 2. We can see that MAE is different between the Carbon-up and Carbon-down ferroelectric states, although these energy differences also dependents on the values of $U$ and $J$ in first-principles calculations. Generally, the carbon-up polarization introduces doped carriers to bilayer $CrI_3$. As a result, for most choices of $U$ and $J$, the MAE energy under the carbon-up configuration has a smaller MAE. This agrees with previous doping calculations of monolayer $CrI_3$.[65]

**Table 2**. Energy difference between magnetic states with out-of-plane spins and in-plane spins of bilayer $CrI_3$ in the 2L-$CrI_3$/$Sc_2CO_2$ heterostructure calculated by different values of $U$ and $J$.

| MAE (meV/unit-cell) | U = 3.9 eV J = 1.1 eV | U = 3.9 eV J = 0.0 eV | U = 2.9 eV J = 0.7 eV | U = 3.0 eV J = 0.0 eV | U = 3.9 eV, J = 1.1eV (fully relaxed) |
|---|---|---|---|---|---|
| **Carbon-down** | -1.13 | -2.69 | -1.62 | -2.62 | -1.18 |
| **Carbon-up** | -0.32 | -3.17 | -1.2 | -2.5 | -1.21 |

At last, we will estimate the working temperature for 2L-CrI$_3$/Sc$_2$CO$_2$ heterostructure. For the ferroelectric Curie temperature, the ferroelectric transition barrier was found to be 0.52 eV per formula unit for Sc$_2$CO$_2$.[19] Thus, the ferroelectric Curie temperature can be up to room temperature. In this sense, the working temperature for the 2L-CrI$_3$/Sc$_2$CO$_2$ heterostructure is determined by the Néel temperature of 2L-CrI$_3$. In our structure, we do not observe significant changes of exchange couplings of CrI$_3$ in this heterostructure. Therefore, we estimate that the Néel temperature is similar to the result of free-standing cases, which is about 41 K observed by experiments.[24] This indicates that the upper limit of the working temperature of multiferroics is around 41 K.

## 3. CONCLUSIONS

We propose artificial multiferroics in a vdW heterostructure formed by bilayer CrI3 and monolayer Sc$_2$CO$_2$. Particularly, an enhanced magnetoelectric effect is predicted: the switch of electronic polarization in the FE Sc$_2$CO$_2$ layer can switch the interlayer magnetic coupling of bilayer CrI$_3$, switching the interlayer FM/AFM magnetic ordering of bilayer CrI3. We further show that the strong build-in electric field in the Sc$_2$CO$_2$ layer alters the energy band alignment of the heterostructure between type II and type III. In the type III band alignment, the charge transfer between CrI$_3$ and Sc$_2$CO$_2$ converts the interlayer magnetic coupling from AFM to FM. Our finding provides a general method to construct 2D artificial-multiferroic devices by stacking of 2D ferroic vdW materials, shedding light on realizing multifunctional nanoelectronic devices.

## 4. COMPUTATIONAL METHOD

Calculations are performed by using first-principles density functional theory (DFT) as implemented in *Vienna Ab initio Simulation Package* (VASP),[66,67] with the dispersion-corrected PBEsol[68] and HSE[63] functional. The first Brillouin zone is sampled with the 9×9×1 and 6×6×1 k-point meshes for DFT and HSE calculations, respectively. A plane-wave basis set with a kinetic energy cutoff of 450 eV is adopted throughout these calculations. A vacuum distance larger than 15 Å is set between adjacent slabs to avoid spurious interactions. The main data are calculated by the GGA+U functional based on the Liechtenstein scheme[69] with $U$ = 3.9 eV and $J$ = 1.1 eV. Other values of $U$ and $J$ have also been tested and give qualitatively similar results. Dipole corrections and SOC are included in all calculations. The differential charge distributions are visualized by the VESTA software.[70]

For structure relaxations, all atoms are relaxed until the residual force per atom is less than 0.005 eV/Å. Because of the lattice mismatch, we fix the experimental in-plane lattice constant of $CrI_3$ and relax all other degree of freedom of the heterostructure. Meanwhile, we have calculated the fully relaxed structure and do not find any qualitatively change of our results. (see the last column of Table 1) Moreover, as shown in Figure 1a, the lowest-energy stacking configuration of bilayer $CrI_3$ and $Sc_2CO_2$ is that the middle-layer Cr atoms are on top of O and C atoms of the bottom $Sc_2CO_2$ layer.

Other interlayer configurations have been checked, and they do not qualitatively change the results.

## AUTHOR INFORMATION

**Corresponding Authors**

*E-mail: liwang@ncu.edu.cn (L.W.)

*E-mail: lyang@physics.wustl.edu (L.Y.)

**Notes**

The authors declare no competing financial interest.

## ACKNOWLEDGEMENTS

Y.L. is supported by the Natural Science Foundation of China (Grant No. 11504158) and China Scholarship Council. L.Y., R.F., X.L., and L.Z. are supported by the National Science Foundation (NSF) CAREER Grant No. DMR-1455346 and the Air Force Office of Scientific Research (AFOSR) grant No. FA9550-17-1-0304. The computational resources have been provided by the Stampede of Teragrid at the Texas Advanced Computing Center (TACC) through XSEDE.

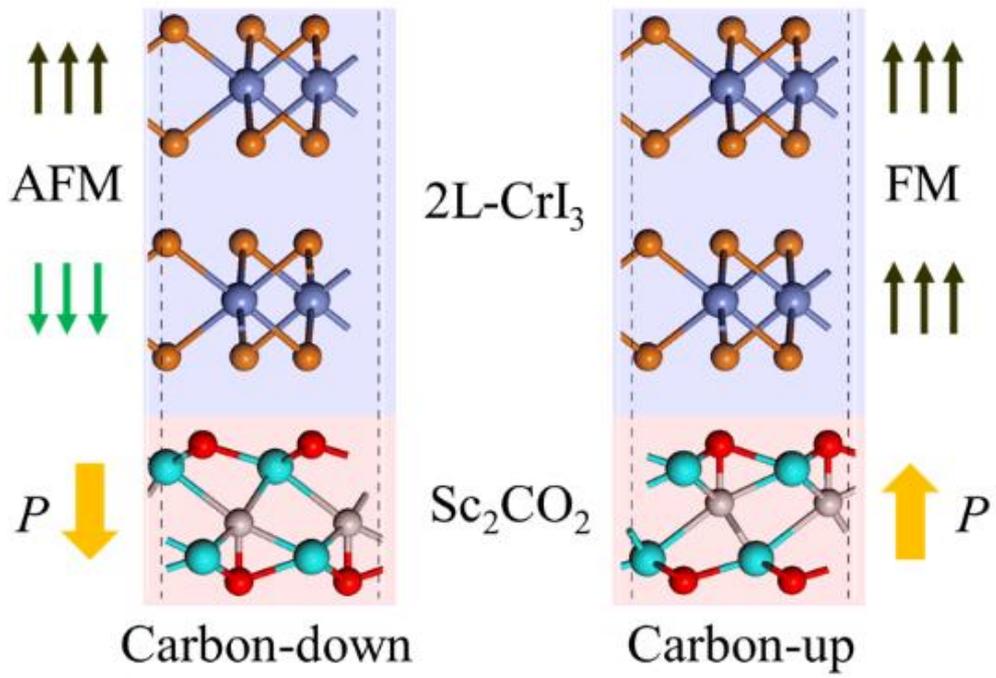

For Table of Contents Only